# Perpendicular magnetic anisotropy in as-deposited CoFeB/MgO thin films

Kaihua Lou[1, 2], Tunan Xie[1, 2], Qianwen Zhao[1, 2], Baiqing Jiang[1, 2], ChaoChao Xia[1, 3], Hanying Zhang[1, 2], Zhihong Yao[1], and Chong Bi[1, 2, 3*]

[1]Key Laboratory of Microelectronic Devices & Integrated Technology, Institute of Microelectronics, Chinese Academy of Sciences, Beijing 100029, China

[2]University of Chinese Academy of Sciences, Beijing 100049, China

[3]School of Microelectronics, University of Science and Technology of China, Hefei 230026, China

*bichong@ime.ac.cn

**Abstract**

Fabrication of perpendicularly magnetized ferromagnetic films on various buffer layers, especially on numerous newly discovered spin-orbit torque (SOT) materials to construct energy-efficient spin-orbitronic devices, is a long-standing challenge. Even for the widely used CoFeB/MgO structures, perpendicular magnetic anisotropy (PMA) can only be established on limited buffer layers through post-annealing above 300 °C. Here, we report that the PMA of CoFeB/MgO films can be established reliably on various buffer layers in the absence of post-annealing. Further results show that precise control of MgO thickness, which determines oxygen diffusion in the underneath CoFeB layer, is the key to obtaining the as-deposited PMA. Interestingly, contrary to previous understanding, post-annealing does not influence the well-established as-deposited PMA significantly but indeed enhances unsaturated PMA with a thick MgO layer by modulating oxygen distributions, rather than crystallinity or Co- and Fe-O bonding. Moreover, our results indicate that oxygen diffusion also plays a critical role in the PMA degradation at high temperature. These results provide a practical approach to build spin-orbitronic devices based on various high-efficient SOT materials.

Perpendicularly magnetized ferromagnets (FMs) have been widely adopted in modern spintronic devices, such as high-density non-volatile magnetic memory, because they can address both thermal stability and power consumption issues simultaneously as the minimum device feature size scales down to several tens of nanometers[1,2]. Typically, there are two types of FMs showing perpendicular magnetic anisotropy (PMA): one is the ferromagnetic materials where PMA persists in a large thickness range, for example, Gd- or Tb-based ferromagnetic alloys with bulk PMA and crystallized Co/Ni or Pt/Co multilayers[2–4]. These materials usually accompany strong PMA and high thermal stability but require huge energy consumption for electrical manipulation of magnetization; the other one is the ultrathin ferromagnetic single layer (less than 2 nm), initially mentioned in the Néel's pioneering work[5], where out-of-plane magnetization is stabilized through interfacial PMA that overcomes the demagnetization field[1,6,7]. In this type of material, the ultrathin FM is usually sandwiched between two heavy-metal (HM) layers[7] or a HM and an oxide layer[1,6,8]. The representative structures are Pt/Co/Pt[7], Pt/Co/AlO$_x$[6] and HM/CoFeB/MgO[1,8,9] trilayers, in which particular interests are focused on the HM/CoFeB/MgO structures since they are compatible with magnetic tunnel junctions (MTJs) producing giant tunneling magnetoresistance[1] and the corresponding magnetization can be easily switched by current-induced spin torques[10–12]. In the HM/CoFeB/MgO structures, it is suggested that the interfacial PMA originates from hybridization of 3$d$ orbitals of Co or Fe and 2$p$ orbitals of oxygen at CoFeB/MgO interfaces[1,13]. However, many works do show that PMA also strongly depends on the HM layers even the roles of HM layers on the contribution of PMA are still under debates[10,14–16]. *Liu et al.* suggested that HM layers actually act as a buffer layer to enhance PMA and PMA is solely from CoFeB/MgO interfaces[9,17], while *Peng et al.* demonstrated that the $p$ orbitals of HMs at HM/CoFeB interfaces also make significant contributions to PMA[18]. In a similar HM/FM/GdO$_x$ structure, it turns out that both the HM/FM and FM/GdO$_x$ interfaces can be controlled separately by an applied voltage and PMA originating from both interfaces can be clearly determined in the same device, providing clear evidence of PMA contributions from both interfaces[19].

Experimentally, the as-deposited CoFeB/MgO structures usually show in-plane anisotropy and an essential annealing process around 300 °C is required to induce PMA[9,15–17,20–26]. The post-deposition annealing is believed to improve the crystallinity of both MgO and CoFeB layers, forming a bcc CoFe/MgO (100) interface[27,28] and improving Fe-O and Co-O bonding[29], so that a large PMA as predicted by the first-principles calculation[13] can be achieved. Moreover, the annealing temperature and time must be carefully controlled per different HMs in case PMA degrades caused by over-annealing[20–22,27]. For instance, PMA of Ta/CoFeB/MgO structures begins to degrade from 300 °C annealing and cannot tolerate the back end of line (BEOL) temperature of 400 °C due to Ta migrations toward the CoFeB/MgO interfaces[9,20,22,27,29]. One interesting question is if PMA of CoFeB/MgO structures can be established reliably in as-deposited samples in the absence of post-annealing processes. This will not only provide additional experimental results to verify PMA mechanisms but also be a great advantage to construct various spintronic devices, especially for the spin-orbitronic devices based on rapidly developing quantum materials (QMs) showing large spin-orbit torques (SOTs)[30–40]. This is because, to utilize QMs in MTJ-compatible devices, another HM insertion layer is usually introduced between the QM and CoFeB layers to create PMA[31,32,40], in which quantum surface states contributing SOTs may be hindered and further destroyed by interfacial mixing during post-annealing processes. So far, growth of CoFeB/MgO structures with as-deposited PMA still remains elusive and there are no reports specifying the means of creating as-deposited PMA in the CoFeB/MgO systems reliably, even though a few works mentioned that PMA could be observed in as-deposited samples[14,28,41].

Here we systematically investigate the possible as-deposited PMA in various CoFeB/MgO structures and demonstrate that the perpendicularly magnetized CoFeB layer can be grown reliably on different buffer layers by precisely controlling subsequently deposited MgO thicknesses. To reveal underlying mechanisms of as-deposited PMA, the crystalline structures and element distributions of as-deposited samples with different MgO thicknesses were further examined by using high-resolution transmission electron microscopy (HRTEM) and scanning transmission electron microscopy (STEM) equipped with energy dispersive X-ray spectroscopy (EDS). The CoFeB/MgO structures with a full stack of buffer layer/Co$_{20}$Fe$_{60}$B$_{20}$ ($t_{CoFeB}$)/MgO ($t_{MgO}$)/Ta (3) were deposited on silicon wafers with 300 nm thermally oxidized SiO$_2$ by using magnetron sputtering, where the numbers or symbols in parentheses indicate thickness in nm and the top 3 nm Ta layer would be naturally oxidized as a capping layer. Two representative HMs (Ta and W) and a typical antiferromagnet (IrMn) were chosen as the buffer layer to investigate the possible as-deposited PMA. The base pressure before sputtering was better than $3 \times 10^{-8}$ Torr. All metallic layers were deposited by using DC magnetron sputtering under an Ar pressure of 2 mTorr and the MgO layer was deposited by using RF magnetron sputtering with a pressure of 1 mTorr. All samples were patterned into a Hall bar structure with the width of 2-10 μm by using standard photolithography and Ar ion milling. The annealing process was performed on the top surface of a hotplate in a glovebox under N$_2$ atmosphere. For electrical measurements, the current was provided by a Keithley 6221 current source and the voltage was monitored by a Keithley 2000 multimeter. All results presented in this work are measured from as-deposited samples except those specifying detailed annealing conditions.

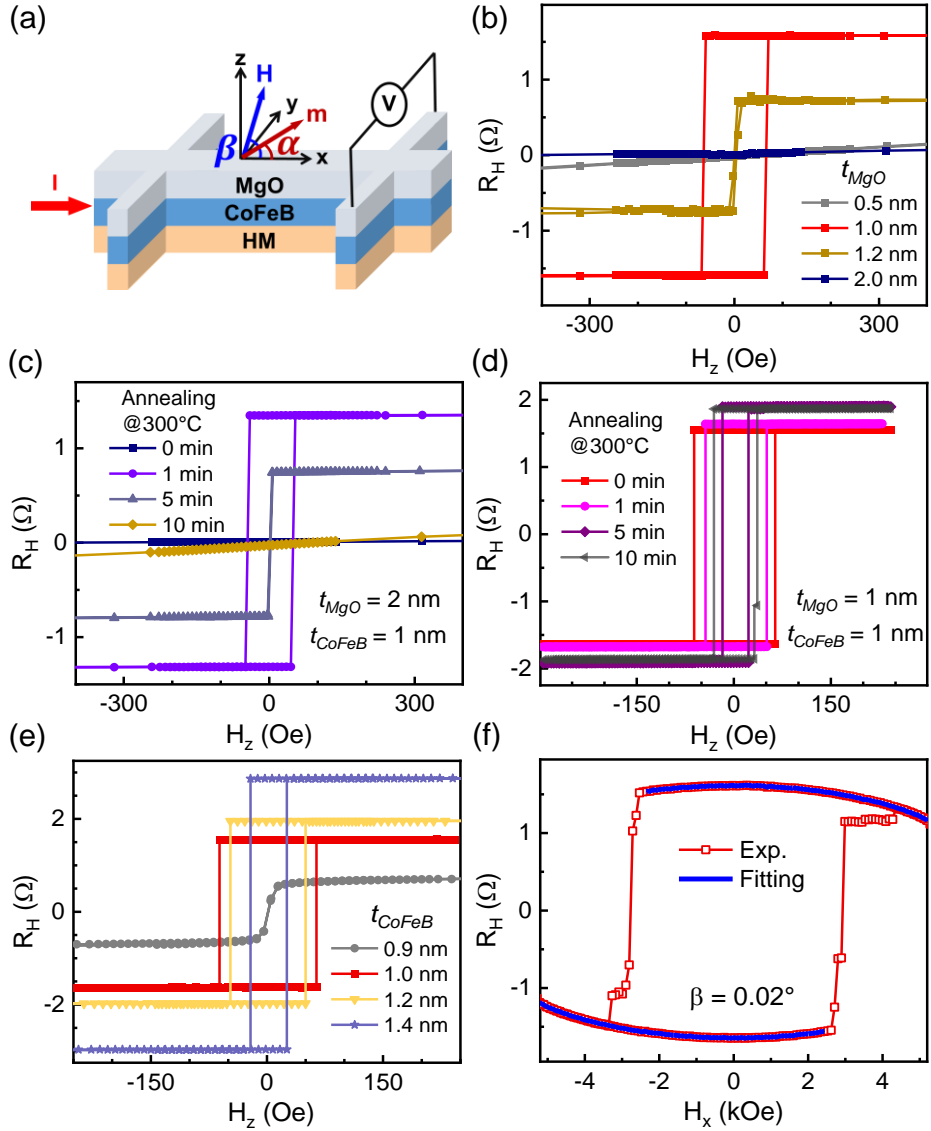

FIG. 1. (a) Schematic of sample structure and experimental configuration. (b) $R_H$ as a function of $H_z$ for as-deposited Ta (8)/CoFeB (1)/MgO ($t_{MgO}$)/TaO$_x$ stacks with various $t_{MgO}$. Comparison of perpendicular $R_H$ curves for the as-deposited and annealed Ta (8)/CoFeB (1)/MgO ($t_{MgO}$)/TaO$_x$ stacks with (c) $t_{MgO}$ = 2 nm and (d) 1 nm. The annealing process was performed at 300 °C for 1 - 10 min. (e) $R_H$ as a function of $H_z$ for as-deposited Ta (8)/CoFeB ($t_{CoFeB}$)/MgO (1)/TaO$_x$ stacks with various $t_{CoFeB}$. (f) Typical in-plane field dependent $R_H$ curves for the Ta (8)/CoFeB (1)/MgO (1)/TaO$_x$ stacks. The in-plane field was tilted about β = 0.02° to prevent domain nucleation and solid lines are fitting results by using the Stoner-Wohlfarth model.

As schematically shown in Fig. 1(a), PMA and magnetization switching were investigated by utilizing the anomalous Hall effect, which has been proven to be an effective approach to investigate perpendicularly magnetized systems since the anomalous Hall resistance ($R_H$) is proportional to the perpendicular component of magnetization ($M_z$)[42]. We first chose Ta as the buffer layer to explore possible as-deposited PMA by optimizing the CoFeB and MgO layers and then demonstrated that the as-deposited PMA can be extrapolated to other buffer layers. Fig. 1(b) presents the recorded $R_H$ as a function of applied perpendicular magnetic field ($H_z$) for Ta (8)/CoFeB (1)/MgO ($t_{MgO}$)/TaO$_x$ stacks, where the sharp magnetization switching and a clear square switching loop can be observed when $t_{MgO}$ = 1 nm, indicating that PMA can be established in the as-deposited samples. When $t_{MgO}$ = 0.5 nm, no magnetization switching occurs, which can be understood that the MgO layer is too thin to form enough Fe-O and Co-O bonds for creating PMA at the CoFeB/MgO interface, and thus, the CoFeB layer shows in-plane magnetic anisotropy. Interestingly, for a thicker MgO layer, the switching loop shrinks ($t_{MgO}$ = 1.2 nm) and even disappears ($t_{MgO}$ = 2 nm). This is counterintuitive but consistent with most works reporting no as-deposited PMA, in which $t_{MgO}$ was usually larger than 1 nm [9,15–17,20–26]. Generally, if PMA has been well established at the CoFeB/MgO interface when $t_{MgO}$ = 1 nm, it should become stronger or at least keeps constant for a thicker MgO layer. The fact that the as-deposited PMA is destroyed by a thicker MgO layer shown in Fig. 1(b) indicates that the subsequently sputtered extra MgO after 1 nm alters the CoFeB/MgO interface or even CoFeB layer. Moreover, the amplitude of $R_H$ for $t_{MgO}$ = 1.2 nm becomes much smaller than that for $t_{MgO}$ = 1 nm, and thus, it is reasonable to assume that a thick MgO layer with $t_{MgO}$ > 1 nm introduces excessive oxygen ions ($O^{2-}$) in the underneath layers breaking the as-deposited PMA and the CoFeB layer has been partially oxidized. This is also reflected in the $R_H$ curve of $t_{MgO}$ = 2 nm, where $R_H$ signal is much weaker compared to $t_{MgO}$ = 0.5 nm, probably because most of the CoFeB layer has been oxidized. It should be noted that interfacial stress contributing PMA in the crystallized Pt (Pd)/Co multilayers due to lattice mismatch[4] can be ignored in the as-deposited CoFeB/MgO structures since both the HM and CoFeB layers are amorphous as confirmed by following HRTEM results.

The surprising results are shown in Fig. 1(c), where the samples with $t_{MgO}$ = 2 nm become perpendicularly magnetized after annealing at 300 °C for 1 minute. This indicates that the annealing process induces $O^{2-}$ redistribution and reduces the CoFeB layer from its oxide counterpart formed in the as-deposited state, even though the reduction behaviors are not expected since no reductant or electrical field is applied[43]. Nonetheless, the reduction of CoFeB was indeed observed by using x-ray photoemission spectroscopy in the initial stage of annealing process [44]. The recovery of PMA due to $O^{2-}$ redistribution and CoFeB reduction is sharply in contrast to previous understanding that the PMA establishment during annealing process is mainly caused by improving crystallinity of CoFeB/MgO interfaces or enhancing Co-O and Fe-O bonding[27–29,41]. As discussed below, this explanation is also supported by quantitative analyses of PMA after post-annealing. Another remarkable feature of Fig. 1(c) is the disappearance of PMA after 10-min annealing which is the well-known problem of Ta/CoFeB/MgO systems that PMA cannot tolerate BEOL temperature attributing to Ta diffusion [9,20,22,27,29]. However, Fig. 1(d) shows that PMA still keeps well for a thinner MgO layer of $t_{MgO}$ = 1 nm after the 10-min annealing, which indicates that, in addition to the Ta diffusion, excessive $O^{2-}$ diffusion and resultant CoFeB oxidization also play critical roles in the PMA degradation at high temperature. Fig. 1(e) shows the evolution of $R_H$ curves by modulating CoFeB thickness while keeping $t_{MgO}$ = 1 nm. It shows that the as-deposited PMA can be established for the CoFeB layer with $t_{CoFeB}$ up to 1.4 nm. Similar to $t_{MgO}$ = 0.5 nm, the weak PMA of $t_{CoFeB}$ = 0.9 nm can also be understood because there are no sufficient Fe-O and Co-O bonds formed at the CoFeB/MgO interface or reduced magnetization due to a dead layer at the Ta/CoFeB interface. All these results indicate that control of $t_{MgO}$ is the key to obtaining as-deposited PMA and there is an annealing window for $O^{2-}$ redistribution and CoFeB reduction to create PMA in the samples with a thick MgO layer. These results explain why annealing is an essential process for PMA establishment in most cases with a thick MgO layer [9,15–17,20–26].

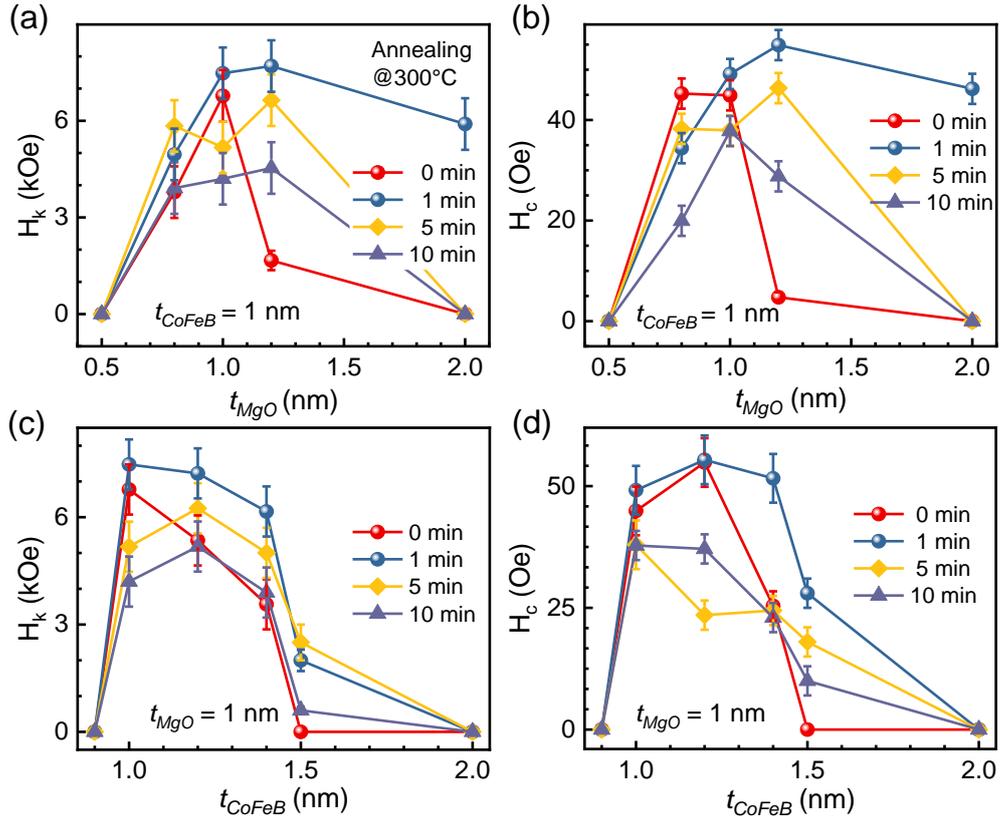

FIG. 2. Annealing effects on extracted (a) $H_k$ and (b) $H_c$ of Ta (8)/CoFeB (1)/MgO ($t_{MgO}$)/TaO$_x$ stacks as a function of $t_{MgO}$. (c) $H_k$ and (d) $H_c$ of Ta (8)/CoFeB ($t_{CoFeB}$)/MgO (1)/TaO$_x$ stacks as a function of $t_{CoFeB}$.

To quantitatively evaluate PMA, $R_H$ as a function of in-plane magnetic field ($H_x$) was measured to estimate PMA field ($H_k$). As shown in Fig. 1(f), the in-plane $R_H$ curve can be fitted by using the Stoner-Wohlfarth model, from which $H_k$ can be extracted[19,45]. Fig. 2 shows the evolution of extracted $H_k$ and coercive field ($H_c$) of Ta (8)/CoFeB ($t_{CoFeB}$)/MgO ($t_{MgO}$)/TaO$_x$ stacks annealed at 300 °C. As shown in Fig. 2(a), the as-deposited PMA for $t_{CoFeB}$ = 1 nm can be established when 0.8 nm ≤ $t_{MgO}$ ≤ 1.2 nm with a maximum value of $H_k$ ≈ 7 kOe, which is comparable to the reported PMA enhanced by post-annealing [9,15–17,20–26]. Moreover, for the samples with a thin MgO layer (0.8 nm ≤ $t_{MgO}$ ≤ 1 nm), as analyzed above that there is no excessive O$^{2-}$ near the CoFeB layer, annealing does not promote $H_k$ significantly even the crystallization of CoFeB/MgO interfaces may be improved. When $t_{MgO}$ ≥ 1.2 nm, the as-deposited PMA begins to degrade due to the appearance of excessive O$^{2-}$, annealing does increase $H_k$ by modulating O$^{2-}$ distribution. When $t_{MgO}$ = 2 nm, PMA is quenched more quickly (less than 5 min) than that for $t_{MgO} \leq 1.2$ nm with increasing annealing time because of further oxidization induced by diffused $O^{2-}$ from the MgO layer. As shown in Fig. 2(b), $H_c$ shows similar behaviors as $H_k$ during annealing. These results further confirm that PMA influenced by annealing is mainly through $O^{2-}$ modulation rather than Co-O or Fe-O bonding and crystallinity of CoFeB/MgO interfaces[27–29,41]. Fig. 2(c) and 2(d) show that the as-deposited PMA for $t_{MgO} = 1$ nm can be established when $1.0$ nm $\leq t_{CoFeB} \leq 1.4$ nm and the annealing effects on PMA do not show strong dependence on $t_{CoFeB}$, which also agrees with our understanding that $O^{2-}$ modulation during annealing mainly occurs near the CoFeB/MgO interface.

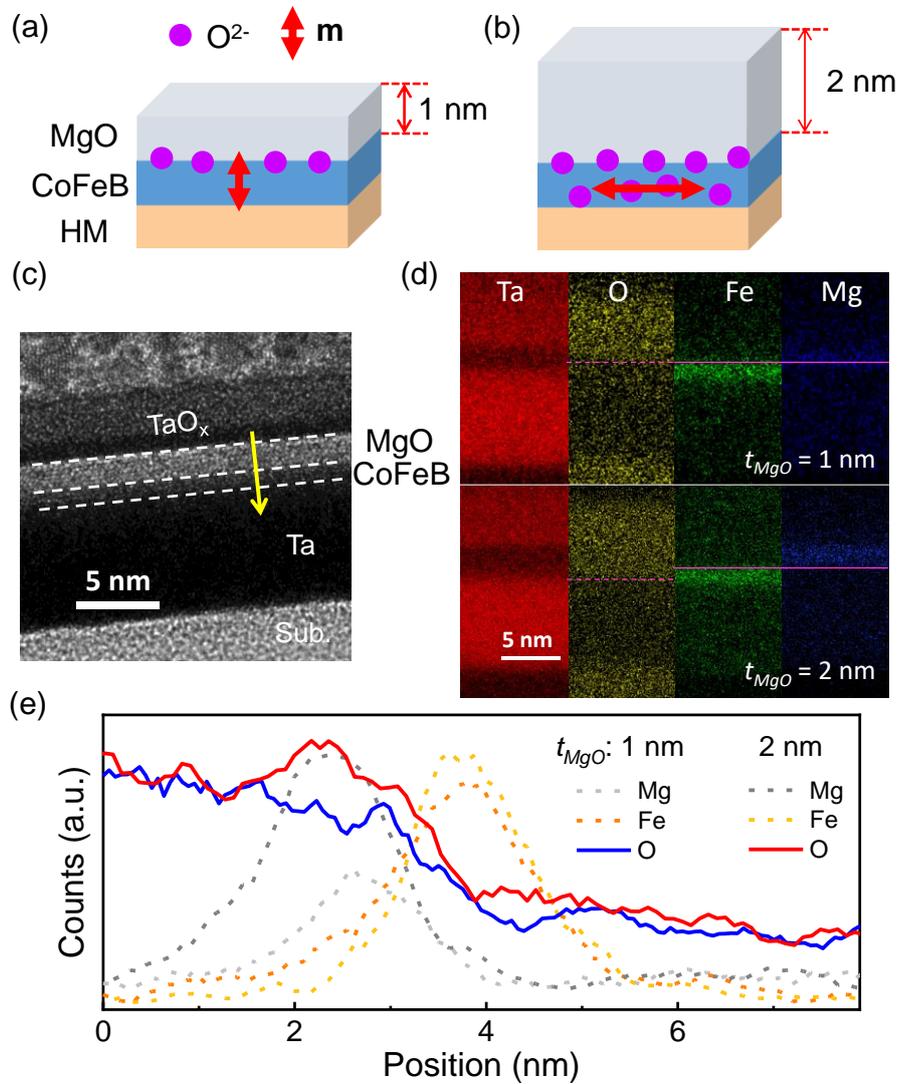

FIG. 3. Schematic of oxygen distribution and magnetization for as-deposited HM/CoFeB/MgO stacks with (a) $t_{MgO} = 1$ nm and (b) 2 nm. (c) Typical HRTEM image of as-deposited Ta (8)/CoFeB (1)/MgO (2)/TaO$_x$ stacks. (d) Ta, O, Fe, and Mg elemental maps of Ta (8)/CoFeB (1)/MgO ($t_{MgO}$)/TaO$_x$ stacks with $t_{MgO}$ = 1 nm (top) and 2 nm (bottom). Solid lines indicate CoFeB/MgO interfaces and dash lines indicate the top boundary of oxygen. (e) Corresponding line profiles of element Mg, Fe, and O for both $t_{MgO}$ = 1 nm and 2 nm along the marked direction in (c).

To clearly clarify the mechanism of as-deposited PMA, O$^{2-}$ distributions in HM/CoFeB/MgO ($t_{MgO}$) stacks for $t_{MgO}$ = 1 nm showing as-deposited PMA and $t_{MgO}$ = 2 nm without PMA are schematically illustrated in Fig. 3(a) and 3(b), respectively. As explained above, PMA arises for $t_{MgO}$ = 1 nm because the Co-O and Fe-O bonds have been formed in the as-deposited state, while it is absent for $t_{MgO}$ = 2 nm due to the over-oxidation of CoFeB layer induced by diffused O$^{2-}$. This is further confirmed by HRTEM and EDS results. Fig. 3(c) shows a typical HRTEM image of as-deposited Ta (8)/CoFeB (1)/MgO (2)/TaO$_x$ stacks, which implies that the Ta and CoFeB layers are amorphous while the MgO layer has been partially crystallized during deposition. Fig. 3(d) presents the elemental mapping images of as-deposited Ta (8)/CoFeB (1)/MgO ($t_{MgO}$)/TaO$_x$ samples with $t_{MgO}$ = 1 nm (top) and 2 nm (bottom). It shows that the top oxygen boundary stops at the edge between Fe and Mg mapping for $t_{MgO}$ = 1 nm, but exceeds the Fe/Mg edge significantly for $t_{MgO}$ = 2 nm, which directly confirms that oxygen has penetrated into the CoFeB layer for thick MgO layers. Fig. 3(e) delineates the comparison of elemental line scanning profiles for both $t_{MgO}$ = 1 nm and 2 nm along the arrow direction in Fig. 3(c). The Fe and right side of Mg signals are aligned for both samples to determine the relative position of O profiles. One can see that the edge of O signal for $t_{MgO}$ = 2 nm is much closer to the Fe signals than that for $t_{MgO}$ = 1 nm, consistent with the schematic mechanisms shown in Fig. 3(a) and 3(b).

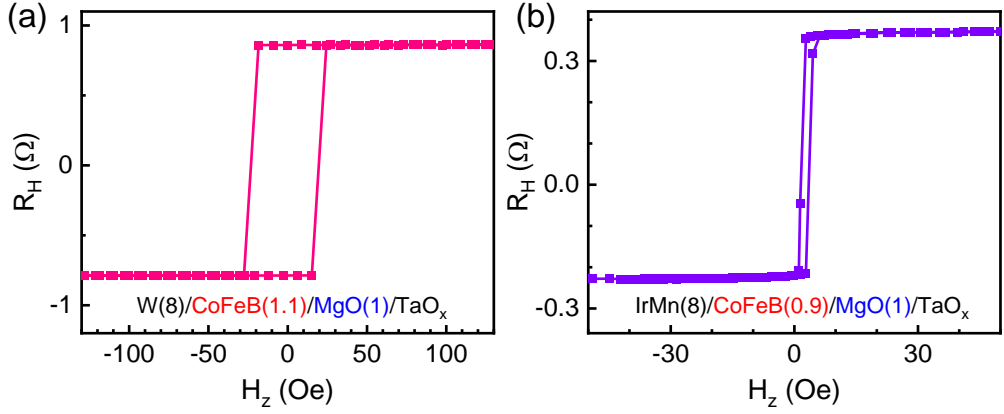

FIG. 4. $R_H$ as a function of $H_z$ for as-deposited (a) W (8)/CoFeB (1.1)/MgO (1)/TaO$_x$ and (b) IrMn (8)/CoFeB (0.9)/MgO (1)/TaO$_x$ stacks. The shift of $R_H$ loop towards positive $H_z$ in (b) indicates that a tilted exchange bias is also induced in the as-deposited state even no external magnetic fields applied during sputtering.

To explore the possible role of buffer layers on the as-deposited PMA, CoFeB/MgO structures with $t_{MgO}$ = 1 nm were also deposited on other buffer layers. Since $t_{MgO}$, and thus the oxygen distribution near the CoFeB/MgO interface, keeps the same for all samples, the variation of as-deposited PMA mainly arises from the influence of buffer layers. By modulating $t_{CoFeB}$, we demonstrate that the as-deposited PMA can also be established reliably on other buffer layers such as W and IrMn. However, the lower limit of $t_{CoFeB}$ for observing as-deposited PMA varies, which is 1.1 nm and 0.9 nm for W (Fig. 4(a)) and IrMn (Fig. 4(b)) buffer layers, respectively. The slight difference of lower limits of $t_{CoFeB}$ among different buffer layers (1.0 nm for Ta, as shown in Fig. 2(c)) may be caused by different dead layers at the buffer layer/CoFeB interfaces. This indicates that $t_{CoFeB}$ must also be modulated accordingly for different buffer layers to obtain as-deposited PMA. Moreover, the as-deposited PMA for Ta and W buffer layers is much stronger than that for IrMn, which reflects the PMA contribution from buffer layer/CoFeB interfaces. Since the as-deposited PMA can be established even in an antiferromagnetic IrMn buffer layer usually showing in-plane exchange coupling[25,26,46] (the shift of $R_H$ loop towards positive $H_z$ shown in Fig. 4(b) is probably due to a tilted in-plane exchange bias), it should be expected by using other non-magnetic SOT materials as the buffer layer.

In summary, we demonstrate that the perpendicularly magnetized CoFeB/MgO structures can be deposited on various buffer layers, in the absence of annealing processes, by controlling MgO and CoFeB thicknesses precisely. Our results show that the as-deposited PMA can only be established when 0.8 nm ≤ $t_{MgO}$ ≤ 1.2 nm and a thick MgO layer ($t_{MgO}$ > 1 nm) will cause the degradation of as-deposited PMA because of oxygen diffusion in the underneath CoFeB layer, as directly confirmed by HRTEM and EDS results. Moreover, for the samples with strong as-deposited PMA (0.8 nm ≤ $t_{MgO}$ ≤ 1.0 nm), annealing process does not influence PMA significantly, while for the samples with unsaturated PMA (1.2 nm ≤ $t_{MgO}$ ≤ 2.0 nm), PMA can be enhanced by modulating oxygen distribution through post-annealing. We hope that the as-deposited PMA can also be established by using other buffer layers, especially the rapidly developing QMs, to build high-efficient SOT devices.

**Acknowledgments**

This work is supported by the National Key R&D Program of China (Grant No. 2019YFB2005800 and 2018YFA0701500), the National Natural Science Foundation of China (Grant No. 61974160, 61821091, and 61888102), and the Strategic Priority Research Program of the Chinese Academy of Sciences (Grant No. XDB44000000).

**Competing interests**

The authors declare no competing interests.

**Data availability**

The data that support the findings of this study are available from the corresponding author upon reasonable request.